\documentclass{article}
\usepackage{amsmath}
\usepackage{amssymb}
\usepackage{mathtools}

\usepackage{amsthm}

\usepackage{amsfonts}

\usepackage{stackengine}

\usepackage{cite}

\DeclarePairedDelimiterX{\inp}[2]{\langle}{\rangle}{#1, #2}
\title{Symplectic Transformations on Wigner Distributions and Time Frequency Signal Design}
\author{Eren Berk Kama \and Mustafa Kuzuoğlu}
\date{}
\begin{document}
\maketitle
\begin{abstract}
This work considers uncertainty relations on time frequency distributions from a signal processing viewpoint. An uncertainty relation on the marginalizable time frequency distributions is given. A result from quantum mechanics is used on Wigner distributions and marginalizable time frequency distributions to investigate the change in variance of time and frequency variables from a signal processing perspective. Moreover, operations on signals which leave uncertainty relations unchanged are studied.
\end{abstract}

\section{Introduction}

Time frequency distributions (TFDs)  are used in many signal processing applications such as in telecommunications, radar, sonar, audio signal processing and biomedical areas. They are used for nonstationary signal analysis, blind source separation and time varying filter design. Examples of applications can be found in \cite{boashash2015time}. A very important issue in time frequency analysis is the uncertainty relations between time and frequency variables. It puts a constraint on how well a signal and its Fourier transform counterpart can be localized in the phase space ((t,f) domain). Studies on harmonic analysis bring useful tools to time frequency signal processing.
\newline
There are different types of uncertainty relations in the literature, a variety of examples are surveyed in \cite{Folland_1997}, \cite{ricaud2014survey}. Uncertainty relations related with time frequency analysis can be found in \cite{grochenig2001foundations},\cite{binz2008geometry}. Implications of these relations on time frequency signal processing are given in \cite{boashash2015time}. We are interested in the relations regarding time frequency distributions. As other TFDs can be obtained by filtering the Wigner distribution, we will be mainly working on it with relations to some marginalizable TFDs. In \cite{narcowich1990geometry}, \cite{littlejohn1986semiclassical}, \cite{de2011symplectic},\cite{de2017quantum} covariance matrix of Wigner distributions was investigated from a quantum mechanical viewpoint. It was shown that, with the action of a symplectic matrix, a new Wigner distribution is created from an existing one, under certain assumptions. Moreover, \cite{narcowich1990geometry} showed that the covariance matrix of a Wigner distribution transfroms in relation with the same symplectic matrix. This gives an invariance result on uncertainty relation. \cite{schempp1984radar}, \cite{dias2019quantum} investigates some actions of symplectic group on ambiguity functions and Wigner distributions. In \cite{auslander1985radar},\cite{miller2002topics},\cite{moran2001mathematics}   harmonic analysis of ambiguity functions was studied in terms of the representations of the Heisenberg group. In \cite{auslander1985radar}, action of special linear group $SL(2,\mathbb{R})$ on the ambiguity functions was given. Ambiguity functions are related with the Wigner distributions via a symplectic Fourier transform. Here, we will use the action of $SL(2,\mathbb{R})$ on Wigner distributions to see which signals could be used to obtain Wigner distributions with unchanged uncertainty relations.
\newline
The rest of the paper is as follows. Section 2 contains the preliminaries on time frequency distributions, ambiguity functions and uncertainty relations. Section 3 gives some uncertainty relations on TFDs. Section 4 gives the action of $SL(2,\mathbb{R})$ and related properties. Section 5 concludes the paper.

\section{Preliminaries}

\begin{description}
\item[Ambiguity functions and TFDs]
\end{description}
In this section, we will give definitions of TFDs and ambiguity functions . More on the topic can be found in \cite{boashash2015time}. The Wigner Distribution of $u,v \in L_2(\mathbb{R})$ is,

\begin{equation*} 
W (u,v)(t,f)  =  \int_{-\infty}^{+\infty} u(t+ \tau/2) v(t - \tau/2) e^{-j2\pi\nu t} d\tau
\end{equation*}

If $u=v$, this relation is called Wigner Distribution of u. The symmetric ambiguity function for $u,v \in L_2(\mathbb{R})$ is,

\begin{equation*} 
A (u,v)(\tau,\nu)  =  \int_{-\infty}^{+\infty} u(t+ \tau/2) v(t - \tau/2) e^{j2\pi\nu t} dt
\end{equation*}

for the $u=v $ case, this relation is called the self ambiguity function. 

 One can use the analytic signal $a(t) = u(t) + j H \{ u(t) \}$ in Wigner Distributions to avoid negative frequency interference. When $a(t)$ is used the distribution is called Wigner Ville Distribution (WVD) $W_{a}(t,f)$. A general TFD $\rho_{a}(t,f)$ can be defined as

\begin{equation*} 
\rho_{a}(t,f) = \gamma(t,f) \underset{t,f}{**} W_{a}(t,f)
\end{equation*}

where $\gamma(t,f)$ is a 2D filter and $**$ denotes convolution in time and frequency. A TFD has time marginal property if, 

\begin{equation*} 
 \int_{-\infty}^{+\infty} \rho_{a}(t,f) df = \vert a(t) \vert^{2}
\end{equation*}

and it has frequency marginal property if it satisfies,

\begin{equation*} 
 \int_{-\infty}^{+\infty} \rho_{a}(t,f) dt = \vert A(f) \vert^{2}
\end{equation*}

Another important property certain TFDs satisfy is the nonnegativity property. It simply holds if $\rho_{a}(t,f) \geq 0$ for all $t,f$. Another important type of TFD is the spectrogram. The spectrogram is defined as,

\begin{equation*} 
S^{w}_{s}(t,f)  = \vert \int_{-\infty}^{+\infty} s(\tau) w(\tau - t) e^{-j2\pi\nu t} d\tau \vert^{2}
\end{equation*}
The spectrogram satisfies the nonnegativity property, but it does not satisfy the marginal properties. As the TFDs can be written in terms of Wigner distributions by filtering, we use the following relation to realise ambiguity functions as the filter domain for TFD design. Between ambiguity functions and WVD, the following relation exists,
\begin{equation*} 
W_{a}(t,f)  = \mathcal{F}\{  \mathcal{F}^{-1} \{ A_{a}(\tau,\nu)\}\}
\end{equation*}
We can define a filtered ambiguity function $\mathcal{A}(\tau,\nu) $, similar to filtering the WVD,
\begin{equation*} 
\mathcal{A}(\tau,\nu) = g(\tau,\nu) A_{a} (\tau,\nu)
\end{equation*}
Then, we can write,
\begin{equation*} 
\rho_{a}(t,f) = \mathcal{F}\{  \mathcal{F}^{-1} \{ \mathcal{A}(\tau,\nu)\}\}
\end{equation*}

Here, we can see that $**$ convolution in time and frequency became multiplication in delay and Doppler. Because of this, ambiguity domain is useful in designing filters. If the kernel $ g(\tau,\nu)$ is in the form, 

\begin{equation*} 
g(\tau,\nu) = G_{1}(\nu) g_{2}(\tau) 
\end{equation*}

it is called a seperable kernel.
\newline
WVD can be seen as a main TFD to modify and obtain other TFDs. WVD satisfies certain plausible properties. It has time marginal and frequency marginal properties, highest resolution in $(t,f)$ among all TFDs. However, it can take negative values and as it is not a linear transformation (bilinear), cross terms deteriorate its performance. On the other hand, spectrogram is nonnegative and has good cross term suppression. Although, it is sensitive to the window size as it increases resolution in time domain by shorter windows, while sacfrificing from the frequency domain, and vice versa. Cross Wigner distributions have applications in quantum mechanics, such as in \cite{de2012weak},\cite{de2012reconstruction}.

\begin{description}
\item[Special Linear Group $SL(2,\mathbb{R})$]
\end{description}

We will give the definition of the special linear group here. Whose actions will be related with ambiguity functions. In\cite{auslander1985radar} $SL(2,\mathbb{R})$ action on ambiguity functions was given. In Section 4, we will use the three generators of $SL(2,\mathbb{R})$ on Wigner distribution and see their effect on variances of time and frequency variables. The special linear group, $SL(2,\mathbb{R})$  is the group of $2\times 2$ real matrices, with determinant 1. It is a noncompact Lie group. Elements of the group are
\begin{equation*}
S= \begin{bmatrix} 
a & b  \\ 
c & d 
\end{bmatrix}
\end{equation*}
for $a,b,c,d \in \mathbb{R}$. Three generators of this group are 

\begin{equation*}
J = \begin{bmatrix} 
0 & 1  \\ 
-1 & 0 
\end{bmatrix}
\end{equation*}

\begin{equation*}
t(a)= \begin{bmatrix} 
1 & 0  \\ 
a & 1 
\end{bmatrix}
, \quad a \in \mathbb{R} 
\end{equation*}

\begin{equation*}
m(b)= \begin{bmatrix} 
b & 0  \\ 
0 & 1/b 
\end{bmatrix}
, \quad b>0 
\end{equation*}

Next, we will define the symplectic form 

\begin{description}
\item[Symplectic Form]
\end{description}

In time frequency analysis and quantum systems, the phase space $\Gamma$ is defined as $\mathbb{R}^{n} \times \mathbb{R}^{n} \cong \mathbb{R}^{2n}$. For n-dimensional case, we have $z=(q_{1},...,q_{n}:p_{1},...p_{n})$. The symplectic form on phase space is defined as,

\begin{equation*} 
\sigma (z,z') = \sum_{i=1}^{n} q'_{i}p_{i}-q_{i}p'_{i}
\end{equation*}

For signal processing problems on 1-dimension, we can use $z=(t,f)$ and the rest of the definitions change accordingly. We will be using the 1-dimensional case here. A $2 \times 2$ matrix $S$ is symplectic iff,

\begin{equation*} 
\sigma (Sz,Sz') = \sigma (z,z')
\end{equation*}

or, equivalently, 
\begin{equation*} 
S^\intercal J S = J , \; for \; J = \begin{bmatrix} 
0 & 1  \\ 
-1 & 0 
\end{bmatrix}
\end{equation*}

\begin{description}
\item[Covariance Matrix of a Wigner Distribution]
\end{description}

The covariance of a Wigner distribution is directly related with the uncertainty relations, as they are statements on variances of time and frequency variables. That is the main reason we are interested in using them for time frequency signal processing applications. In \cite{narcowich1990geometry}, \cite{littlejohn1986semiclassical}, it was shown that, the covariance matrix of a Wigner distribution has the invariance property under certain linear transformations. We will state some important results from \cite{narcowich1990geometry}, which are useful for our purposes here. We will not go into too much detail here, as they are in quantum mechanics context, we refer the interested reader for proofs to the paper. For a treatment in terms of quantum harmonic analysis with relations to Wigner distributions and ambiguity functions one can refer to \cite{de2017quantum}.
\newline
As we have mentioned, the Wigner distribution holds the marginal properties. Integrating it over the time variable will yield the frequency marginal and vice versa. We will consider normalized Wigner distributions,

\begin{equation*} 
\int \int_{-\infty}^{+\infty} W(t,f) dt df = 1
\end{equation*}

Moreover, we will assume that the integral,

 \begin{equation*}
 \int_{\Gamma} (1+|z|^{2})   W(z) dz < \infty
 \end{equation*}
is finite. This assumption assures that Fourier transform of the Wigner distribution is twice continuously differentiable on the dual of the phase space $\hat{\Gamma}$. We may find the elements of the covariance matrix as,

 \begin{equation*}
C_{ij} = \int (z_{i}-z_{i0})  (z_{j}-z_{j0})  W(z) dz 
 \end{equation*}

Where, $z_{i 0} , z_{j 0}$ are the means of $z_{i}, z_{j}$ respectively. For one dimensional case, the covariance matrix is,

 \begin{equation*}
C =  \begin{bmatrix} 
\Delta t^{2} & \Delta (t,f)  \\
\Delta (f,t) & \Delta f^{2}

\end{bmatrix}            	
\end{equation*}

Where,

 \begin{equation*}
 \Delta t^{2} = \int \int  t^{2} W(t,f) dt df - \left [ \int \int t  W(t,f) dt df \right ]^{2}= <t^{2}>-<t>^{2}
 \end{equation*}

 \begin{equation*}
 \Delta f^{2} = \int \int  f^{2} W(t,f) dt df - \left [ \int \int f W(t,f) dt df \right ]^{2}= <f^{2}>-<f>^{2}
 \end{equation*}

 \begin{equation*}
 \Delta (t,f) = \int \int  tf W(t,f) dt df - \int \int t  W(t,f) dt df \int \int f  W(t,f) dt df
 \end{equation*}

 \begin{equation*}
= <tf>-<t><f>
 \end{equation*}
Here, we can note that time frequency distributions which satisfy the marginality conditions can also be used to obtain the $C$ matrix. The dual of $\rho(t,f)$ is $A_{u}(\tau,\nu) g(\tau,\nu)$. Here, the kernel $g(\tau,\nu)$ must satisfy the conditions for marginality given in section 2. In order to satisfy the twice differentiability condition, we require $g(\tau,\nu)$ to be twice continuously differentiable in addition to $A_{u}(\tau,\nu)$.
\newline
A very important result of \cite{narcowich1990geometry} is, if $W(z)$ holds the above given assumptions then, $C$ is its covariance matrix iff it is real symmetric and the matrix,
 \begin{equation*}
C + \frac {i \hbar}{2} J
 \end{equation*}
is Hermitian, and non-negative. This result is also called the strong uncertainty. From here, by looking at the determinant, we obtain

 \begin{equation*}
\Delta t^{2} \Delta f^{2} \geq \Delta (t,f) + \frac {\hbar^{2}}{4}
 \end{equation*}

Importance of the covariance matrix for our purpose here is that it has an invariance under affine transformations. This relation was deeply investigated in \cite{littlejohn1986semiclassical} and \cite{narcowich1990geometry}. We consider the linear transformation $A : W(z) \rightarrow W(S^{-1}z)$, where S is a symplectic matrix. Here, we omitted the constant part of the affine transformation as we are not going to use it in our analysis. We are interested in the behavior of the covariance matrix under the effect of the symplectic matrix. Simply, by using $S^{-1}z$ in the definition of covariance matrix above, we obtain the following matrix,

 \begin{equation*}
C' = \int [S(z_{i}-z_{i0})]  [S(z_{j}-z_{j0})]^\intercal  W(z) dz 
 \end{equation*}
 \begin{equation*}
= S \left[ \int (z_{i}-z_{i0}) (z_{j}-z_{j0})  W(z) dz \right] S^\intercal = S C S^\intercal
 \end{equation*}
We can see here that $C' + \frac {i \hbar}{2} J$ is also nonnegative, because,

 \begin{equation*}
C' + \frac {i \hbar}{2} J = SC S^\intercal + S (\frac {i \hbar}{2} J )S^\intercal = S \left[ C + \frac {i \hbar}{2} J\right] S ^\intercal
 \end{equation*}
and  we know that $C + \frac {i \hbar}{2} J $ is non negative.

\section{Uncertainty Relations}
Here, we will give relations about ambiguity functions and TFDs. The following relation is the usual Heisenberg uncertainty relation.
\begin{description}
\item[Heisenberg Uncertainty Relation]
\end{description}

Heisenberg uncertainty relation is a well known statement from quantum mechanics. In signal processing context, it means that a signal and its Fourier domain counterpart, cannot be localized in $(t,f)$ domain simultaneously. The uuncertainty relation can be given as follows,

\begin{equation*} 
\int (t-a)^2 \vert x(t) \vert^{2} dt \int (f-b)^2 \vert X(f) \vert^{2} df \geq \frac{\Vert x \Vert_{2}^{4}}{16 \pi^2}
\end{equation*}

for $x \in L_2(\mathbb{R})$. Proof of this relation can be found in \cite{Folland_1997}. In section 2, time marginal and frequency marginal properties were given. Combining these with the uncertainty relations we can use them to prove theorems given for Wigner distributions. Here, by using a marginalizable TFD $\rho_{a}(t,f)$ we will obtain the following uncertainty relation,

\begin{description}
\item[Relation 1]
\end{description}
For a $\rho_{a}(t,f)$ satisfying the time marginal and frequency marginal properties and the constants $t_{0}$ and $f_{0}$,
\begin{equation*} 
 \int  \int \vert t-t_{0}\vert^{2}+\vert f-f_{0}\vert^{2} \rho_{a}(t,f) dt df \geq \frac{\Vert a \Vert_{2}^{2}}{2 \pi}
\end{equation*}

\begin{proof} 
As $\rho_{a}(t,f)$ satisfies marginal conditions,

\begin{equation*} 
 \int_{-\infty}^{+\infty} \rho_{a}(t,f) df = \vert a(t) \vert^{2}
\end{equation*}

and,

\begin{equation*} 
 \int_{-\infty}^{+\infty} \rho_{a}(t,f) dt = \vert A(f) \vert^{2}
\end{equation*}

We have uncertainty relation given in section 2,
\begin{equation*} 
\int (t-a)^2 \vert x(t) \vert^{2} dt \int (f-b)^2 \vert X(f) \vert^{2} df \geq \frac{\Vert x \Vert_{2}^{4}}{16 \pi^2}
\end{equation*}

Also, as $u^{2}+v^{2} \geq 2 uv $, we obtain, 

\begin{equation*}
\int  \int \vert t-t_{0}\vert^{2}+\vert f-f_{0}\vert^{2} \rho_{a}(t,f) dt df =
\end{equation*}

\begin{equation*}
= \int \vert t-t_{0}\vert^{2} \vert a(t)\vert^{2} dt +\int \vert f-f_{0}\vert^{2} \vert A(f)\vert^{2} df  \geq \frac{\Vert a \Vert_{2}^{2}}{2 \pi}
\end{equation*}
\end{proof}

This relation gives an uncertainty for all the TFDs wchich satisfy time marginal and frequency marginal properties. In order to satisfy the time marginal, the TFD's kernel should satisfy $g(0,\nu) = 1, \forall \nu$ and for the frequency marginal $g(\tau,0) = 1, \forall \tau$. Some TFD's satisfying both marginal conditions are, WVD, Levin Distribution, Rihaczek Distribution, Born Jordan Distribution, Gaussian TFD and Page Distributions.
\newline
Gaussian transform of WVD , $W_{\alpha \beta}(t,f)$ can be defined as \cite{de1967uncertainty},

\begin{equation*} 
W_{\alpha \beta}(t,f) = \int  \int_{-\infty}^{\infty}   W_{a}(t,f) e^{-\frac{\pi}{\alpha}  (t-t')^{2}} e^{-\frac{\pi}{\beta}  (f-f')^{2}} dt' df'
\end{equation*}

for $z \in L_2(\mathbb{R})$. $W_{\alpha \beta}(t,f)$ is a filtered version of $W_{a}(t,f)$. If we take $\gamma_{1}(t) = e^{-\frac{\pi}{\alpha} t}$ and $\gamma_{2}(f) = e^{-\frac{\pi}{\beta} f}$, then,
 
\begin{equation*} 
W_{\alpha \beta}(t,f) = \gamma_{1}(t) \underset{t}{*} W_{a}(t,f) \underset{f}{*} \gamma_{2}(f)
\end{equation*}
Actually, $\gamma(t,f)=\gamma_{1}(t)\gamma_{2}(f)$ can be realized as a seperable kernel. A property of Gaussian transformed WVD is that, $W_{\alpha, \beta}(t,f)$ filtered with parameters $\alpha',\beta' $
 gives $W_{\alpha+\alpha', \beta+\beta'}(t,f)$.

The following relation is important as it help to realize $W_{\alpha,\beta}(t,f)$ as a probability distribution. We will give it without the proof. The proof can be found in \cite{de1967uncertainty}.
\newpage

\begin{description}
\item[Relation 2]
\end{description}

Let $\alpha>0$, $\beta>0$. Then, $W_{\alpha,\beta}(a,a)(t,f)$ is positive definite iff $\alpha\beta > 1/4$. 
\newline

As WVD can have negative values, it cannot be considered as a probability distribution. This relation gives $W_{\alpha,\beta}(z,z)(t,f)$ the nonnegativity property, hence, allowing us to consider it as a probability distribution. The relation intuitively means that, if the distribution $W_{\alpha,\beta}(a,a)(t,f)$ is a measure of energy, being nonnegative, then the kernel $\gamma_{1}(t)\gamma_{2}(f)$ has to satisfy $\alpha\beta > 1/4$. As WVD is filtered with Gaussians, this distribution can be interpreted as joint distribution of time to an error $\alpha$ and frequency to an error $\beta$ \cite{Folland_1997}.
\newline
If we look at the ambiguity function counterpart of $W_{\alpha,\beta}(a,a)(t,f)$, and denote it with $A_{\alpha,\beta}(a,a)(t,f)$ we can see from,
\begin{equation*} 
W_{\alpha \beta}(t,f) = \gamma_{1}(t) \underset{t}{*} W_{a}(t,f) \underset{f}{*} \gamma_{2}(f)
\end{equation*}
and,
\begin{equation*} 
\mathcal{A}(\tau,\nu) =  \mathcal{F}\{  \mathcal{F}^{-1} \{ \rho_{a}(t,f)\}\}
\end{equation*}
that,
\begin{equation*} 
\mathcal{A}_{\alpha,\beta}(\tau,\nu) =   A_{a}(\tau,\nu)\mathcal{F}\{ \gamma_{1}(t) \} \mathcal{F}^{-1} \{ \gamma_{2}(f)\}
\end{equation*}
\begin{equation*} 
=   A_{a}(\tau,\nu) G_1(\nu) g_{2}(\tau)
\end{equation*}

As Fourier transform of a Gaussian signal gives a Gaussian. $G_1(\nu)$ and $ g_{2}(\tau)$ are also Gaussians. Here, we see that the effect of filtering the Wigner distribution with Gaussians in the dual domain is multiplication with Gaussians. This effect modifies the variance of the ambiguity distribution in the opposite way. This happens as the Fourier transform of the Gaussian will have the variance inverted.
\newline

\section{$SL(2,\mathbb{R})$ Actions on Wigner Distributions }

In \cite{auslander1985radar}, $SL(2,\mathbb{R})$ action on ambiguity functions were given. Here, we will use them in Wigner distributions and we will also give the ambiguity function counterparts. As we have mentioned earlier, the covariance matrix can be obtained by marginalizable TFDs. Therefore, after applying the group actions on a marginalizable TFD, one can still obtain the covariance matrix. Here, we will only solve for the Wigner distributions.  Our purpose here is to observe the effect of generators of the group $SL(2,\mathbb{R})$. Using relations for generators, we can see the effect of other actions of  $SL(2,\mathbb{R})$, as they can be written in terms of generators. 

\begin{description}
\item[Relation 4.1]
\end{description}
Let $a \in L_2(\mathbb{R}) $, then

\begin{equation*} 
W_{a}(t,f) \circ {J^{-1}} = W_{\mathcal{F}\{a\}}(t,f) = W_{a}(-f,t)
\end{equation*}
\begin{proof} 
The relation can be shown to hold by using the definiton of Wigner distributions and Parseval theorem.
\end{proof} 
If we look at the corresponding ambiguity function, 

\begin{equation*} 
\tilde A_{a}(\tau,\nu) =   \mathcal{F}\{  \mathcal{F}^{-1} \{ W_{a}(-f,t)\}\}  = \mathcal{F}\{  \mathcal{F}^{-1} \{ W_{\mathcal{F}\{a\}}(t,f)\}\}  
\end{equation*}
\begin{equation*} 
= A_{\mathcal{F}\{a\}}(\tau,\nu) = A_{\mathcal{F}\{a\}}(\nu,-\tau)
\end{equation*}
Here, if we look at the covariance matrix,

 \begin{equation*}
C' = J C J^\intercal         	
\end{equation*}
 \begin{equation*}
 = J \begin{bmatrix} 
\Delta t^{2} & \Delta (t,f)  \\
\Delta (f,t) & \Delta f^{2}

\end{bmatrix}   J^\intercal         	
\end{equation*}
 \begin{equation*}
 =  \begin{bmatrix} 
\Delta f^{2} & -\Delta (t,f)  \\
-\Delta (f,t) & \Delta t^{2}

\end{bmatrix}         	
\end{equation*}
We see that, $\Delta t' = \Delta f$, $\Delta f' = \Delta t$ and the uncertainty relation $\Delta t' \Delta f' \geq \frac { \hbar}{2}$ holds.

\begin{description}
\item[Property 4.2]
\end{description}

\begin{equation*} 
(W_{a}(t,f) \circ {J^{-1}} ) \circ {J^{-1}} = W_{a}(-t,-f) = W_{a}^*(t,f)
\end{equation*}

\begin{proof} 
The relation can be proved by applying the operation directly and using the definition of Wigner distributions.
\end{proof} 
The ambiguity function counterpart here is,
\begin{equation*} 
\tilde A_{a}(\tau,\nu) =   \mathcal{F}\{  \mathcal{F}^{-1} \{ W_{a}(-t,-f)\}\}  = \mathcal{F}\{  \mathcal{F}^{-1} \{ W_{a}^*(t,f)\}\}  
\end{equation*}
\begin{equation*} 
= A_{a}(-\tau,-\nu) 
\end{equation*}

The covariance matrix is, 
 \begin{equation*}
C' = (JJ) C (JJ)^\intercal         	
\end{equation*}
As $JJ= (-1) \bf I$,

 \begin{equation*}
C' = (-1)\mathbf {I} C (-1 )\mathbf {I}^\intercal   = C      	
\end{equation*}
The covariance matrix doesnt change. Therefore, $\Delta t' = \Delta t$, $\Delta f' = \Delta f$ and the uncertainty relation
 $\Delta t' \Delta f' \geq \frac { \hbar}{2}$ is unchanged.

\begin{description}
\item[Property 4.3]
\end{description}

\begin{equation*} 
W_{a}(t,f) \circ t^{-1}(k) = W_{a}(t,f - k t)   = W_{a}^{LFM}(t,f)
\end{equation*}

\begin{proof} 
This property can be shown to hold by direct substitution.
\end{proof} 

Here, the effect of $t^{-1}(k)$ changes $a(t)$ to $a_3(t)= a(t) e^{-j\pi k t^{2}}$, which is the LFM effect. Ambiguity function here becomes, 
\begin{equation*} 
A_{a_{3}}(t,f) =   \mathcal{F}\{  \mathcal{F}^{-1} \{ W_{a}(t,f - k t)\}\}  = \mathcal{F}\{  \mathcal{F}^{-1} \{ W_{a}^{LFM}(t,f)\}\}  
\end{equation*}
\begin{equation*} 
= A_{a}(\tau,\nu+k\tau) =A_{a}^{LFM}(\tau,\nu)
\end{equation*}
Which is obtained by using LFM effect in ambiguity functions. We note that, the LFM effect here is given as $(\tau,\nu+k\tau)$, which is usually given as a negative shift, this can be solved by choosing a negative $k$. Here, the corresponding symplectic matrix $S$ is,
 \begin{equation*}
 =  \begin{bmatrix} 
1 & 0  \\
k & 1
\end{bmatrix}         	
\end{equation*}
If we look at the covariance matrix,
 \begin{equation*}
C' = S C S^\intercal    	
\end{equation*}
 \begin{equation*}
 = S \begin{bmatrix} 
\Delta t^{2} & \Delta (t,f)  \\
\Delta (f,t) & \Delta f^{2}

\end{bmatrix}       S^\intercal    	
\end{equation*}

 \begin{equation*}
 =  \begin{bmatrix} 
\Delta t^{2} & \Delta (t,f) +k \Delta f^{2} \\
\Delta (f,t) +k \Delta f^{2}& \Delta f^{2}+2k \Delta (f,t)+k^{2}\Delta t^{2}

\end{bmatrix}          	
\end{equation*}
Here, we have $\Delta t' = \Delta t$, $\Delta f' = \sqrt{\Delta f^{2} +2k \Delta (f,t)+k^{2}\Delta t^{2}}$ and we can easily see that, $\Delta t' \Delta f'  \geq \Delta t \Delta f \geq  \frac { \hbar}{2} $. Therefore, the uncertainty is preserved.

\begin{description}
\item[Property 4.4]
\end{description}
Let $\tilde {a(t)} = a(\tfrac{t}{c})$,  then,

\begin{equation*} 
W_{a}(t,f) \circ m^{-1}(c) = W_{a}(\tfrac{t}{c}, cf)= W_{a_{4}}(\tau,\nu)
\end{equation*}

\begin{proof} 
This property can be proved by directly applying the operation and using the definition of ambiguity functions.
\end{proof} 

$m^{-1}(c)$ operation turns $a(t)$ to $a_4(t)= a(\tfrac{t}{c})$ here. The resulting ambiguity function counterpart is therefore,
\begin{equation*} 
 A_{a_{4}}(t,f) = \tfrac{1}{c}  \mathcal{F}\{  \mathcal{F}^{-1} \{ W_{a}(\tfrac{t}{c}, cf)\}\}  = \mathcal{F}\{  \mathcal{F}^{-1} \{ W_{a_{4}}(t,f)\}\}  
\end{equation*}
\begin{equation*} 
= A_{a_{4}}(t,f) =  \tfrac{1}{c}A_{a}(c\tau,\tfrac{\nu}{c})
\end{equation*}
The symplectic matrix here is,
 \begin{equation*}
S =  \begin{bmatrix} 
c & 0  \\
0 & \tfrac{1}{c}
\end{bmatrix}         	
\end{equation*}
Then, the covariance matrix is
 \begin{equation*}
C' = S C S^\intercal    	
\end{equation*}
 \begin{equation*}
 = S \begin{bmatrix} 
\Delta t^{2} & \Delta (t,f)  \\
\Delta (f,t) & \Delta f^{2}

\end{bmatrix}       S^\intercal    	
\end{equation*}

 \begin{equation*}
 =  \begin{bmatrix} 
c^{2}\Delta t^{2} & \Delta (t,f)  \\
\Delta (f,t) & \tfrac{1}{c^{2}} \Delta f^{2}

\end{bmatrix}          	
\end{equation*}
Therefore, we have $\Delta t' = c \Delta t$, $\Delta f' = \tfrac{1}{c} \Delta f$ and we have, $\Delta t' \Delta f'  = \Delta t \Delta f \geq  \frac { \hbar}{2} $. Therefore, the uncertainty holds. In this section we gave the effect of three generators of $SL(2,\mathbb{R})$ on the Wigner distributions and therefore, on ambiguity functions and marginalizable TFDs. As matrices in $SL(2,\mathbb{R})$ can be written in terms of these three generators, the relations of the others can be deduced from the ones we gave here. We have also investigated which signal corresponds to the resulting Wigner distribution.

\section{Conclusion}
In this paper we have given some uncertainty relations on Wigner distributions and marginalizable TFDs. We used some results in \cite{de1967uncertainty} with a signal processing approach in relation with TFD design. We have used the important relation on Wigner distributions in \cite{narcowich1990geometry}  in the TFD context. We have shown which signals are corresponding to the actions of the special linear group. Moreover, we have investigated the effect of the generators of $SL(2,\mathbb{R})$ on Wigner distributions as symplectic matrices.

\bibliographystyle{plain} 
\bibliography{sympref}

\end{document}